# Are task representations gated in macaque prefrontal cortex?

Timo Flesch[1], Valerio Mante[2], William Newsome[3], Andrew Saxe[4], Christopher Summerfield[1], David Sussillo[3].

[1] University of Oxford
[2] University of Zurich
[3] Stanford University
[4] University College London

Author order is alphabetical. Correspondence: christopher.summerfield@psy.ox.ac.uk, bnewsome@stanford.edu, a.saxe@ucl.ac.uk, valerio@ini.uzh.ch

Abstract

A recent paper (Flesch et al, 2022) describes behavioural and neural data suggesting that task representations are gated in the prefrontal cortex in both humans and macaques. This short note proposes an alternative explanation for the reported results from the macaque data.

Main Text

In a recent paper, Flesch, Saxe and Summerfield (FSS) describe behavioural and neural data recorded whilst human subjects perform a context-dependent decision-making task, and propose a computational model, framed in terms of a neural network, that explains their findings (Flesch et al., 2022). The task required human participants to make, on interleaved trials, one of two category judgments about orthogonal features of a naturalistic image set (trees), with the decision-relevant feature indicated by a contextual cue. They report that the geometry of BOLD signals in neocortical regions (including the prefrontal and parietal cortices) implies that the relevant dimensions for each context are coded along orthogonal dimensions in neural state space, with the irrelevant dimension relatively compressed, which is consistent with other fMRI studies (Ritz and Shenhav, 2022). In their paper, FSS report that this geometry can emerge naturally in a feedforward neural network trained with gradient descent, in particular when the networks is initialised with small weights (the "rich" training regime).

In the same paper, FSS report a re-analysis of single unit data recorded by Mante, Sussillo, Shenoy and Newsome from the frontal eye fields (FEF) of the macaque monkey (Mante et al., 2013). FSS analysed this data because the task was qualitatively similar: monkeys were cued to discriminate two orthogonal features (colour and motion direction) of random dot kinetograms, with the context signalled by a cue that varied from block to block, and behavioural responses made via a saccadic eye movement. FSS analysed these data using the same approach as for the fMRI data: they sorted trials into conditions according to the context and the two stimulus features, and constructed representational dissimilarity matrices (RDMs) that reflected the neural dissimilarity (difference in population activity) between each condition and every other condition. They then adopted an identical analytic approach as for the fMRI data, constructing model RDMs that expressed various coding schemes for these data, including schemes in which the representation of irrelevant stimulus features were either compressed or uncompressed. They found what they took to be a striking correspondence between the monkey data and the results observed in BOLD, and in their paper, report that the monkey data support a "compression" effect similar to that observed in humans. This finding appears at odds with the original analysis of these data (Mante et al, 2013), which concluded that the representations of irrelevant features are largely uncompressed in FEF.

In recent correspondence, Mante Sussillo and Newsome (MSN) proposed a different, more likely explanation for the results FSS obtained from the monkey data. MSN note two differences between how FSS and MSN analysed these data, which impact the respective conclusions about the nature of context-dependent computations in FEF.

First, the analyses by FSS and MSN are based on different task variables (see **Fig. 1**). To appreciate this difference, it is necessary to briefly discuss the task design employed by Mante et al (2013). In their task, the response targets flank the stimulus, e.g. on the left and right, with the colour-side assignments switching randomly from trial to trial. The monkeys are trained to respond to the target whose side matches the motion direction of the moving dots (in the motion context) or whose colour matches the colour of the moving dots (in the colour context). In principle, monkeys (and artificial neural networks) could solve this task in two ways (Fig. 1). Evidence could be accumulated towards a choice (a categorical judgment) in a frame of reference of the inputs (Fig. 1, Flesch coding): to the right or left target in the motion context or to the red or green target in the colour context. Alternatively, stimulus colour could first be transformed into evidence towards the right or left target, and evidence could then be accumulated in the frame of reference of the operant saccade (the response; Fig. 1, Mante coding). While FSS analysed the monkey data in frame of reference of the inputs, MSN analysed it in the frame of reference of the response (see labelled variables in Fig. 1). Beyond incorporating distinct assumptions about the nature of evidence accumulation in the brain, these differences in the respective analyses also preclude a straightforward comparison of the findings in Flesch et al (2022) and Mante et al (2013).

Second, whereas MSN report input signals that are residual (orthogonal) to activity representing the monkey's saccadic response, the analyses of the FEF data by FSS conflate the representations of inputs and response. This conflation results from correlations between the input variables and the response itself. In particular, in the input frame of reference used by FSS, the correlations are asymmetric by design. The motion input is correlated with saccade direction (the response), because the association between motion direction and the position of the correct response target is fixed. The colour input, however, is not correlated with response, because target colour and position are perfectly counterbalanced. When task variables are correlated, activity modulation that appears to be due to one variable may in fact reflect modulation by a correlated variable. It is widely known that many FEF neurons have response fields coding for the direction and amplitude of a forthcoming saccade (Bruce and Goldberg, 1985) and indeed the representation of the upcoming saccade was the dominant signal identified by MSN in the FEF activity (referred to as the "choice" axis in Mante et al, 2013, a different terminology than in this manuscript). To retrieve input representations that were not conflated by these strong response representations, MSN employed an orthogonalization procedure at the level of the neural population activity. Notably, in the BOLD data analysed by FSS (Flesch et al 2022), the inputs are uncorrelated from the response by design, as the categorical choices about the stimulus features are counterbalanced with the responses used to indicate the chosen category. Activity that is modulated by the inputs is then guaranteed to not be contaminated by the response, even without an explicit orthogonalization of the inferred representations, so that the BOLD analyses are unaffected by this issue.

The definition of inputs used by FSS in their analysis of the monkey data, in combination with the conflation of input and response-related activity, can result in spurious evidence for compression of the irrelevant input. In particular, activity representing the monkey's response can be expected to strongly inflate the representation of the motion input when motion is the contextually relevant feature, but not when it is the irrelevant feature, creating an effect akin to a compression of the irrelevant feature. Indeed, simulations of the monkey data based on the mechanism of late selection proposed by Mante et al (2013), which relies on "uncompressed" inputs, reproduce the outcome of the RDM analyses that FSS took as providing evidence towards compression (not shown).

Thus, whilst the study by FSS reports evidence for compression of irrelevant information in BOLD signals, the monkey data is consistent with largely uncompressed input representations, as concluded by MSN in their original analyses of these data (Mante et al, 2013). Future studies will have to address the reasons for the discrepancy between the monkey and BOLD data, which may reflect differences in recording method, stimuli, species, or brain area.

**Figure 1.**

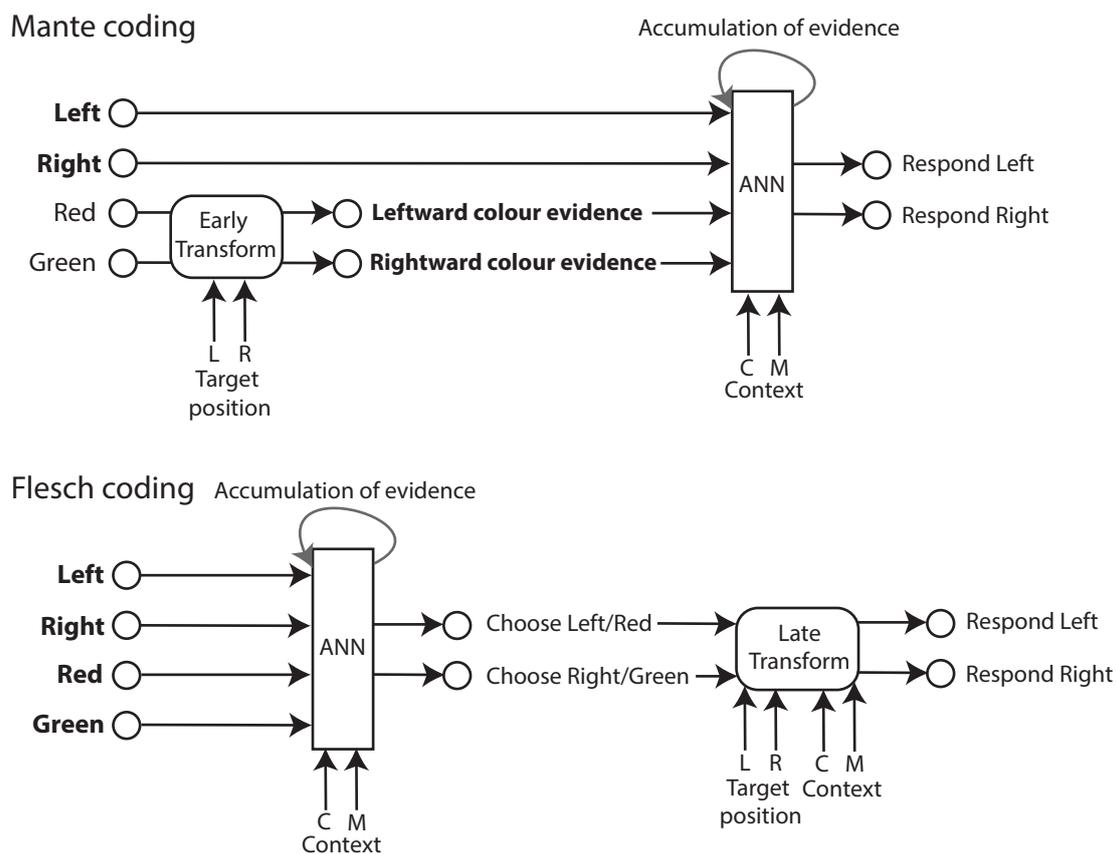

**Figure 1:** Different assumptions about accumulation of evidence in MSN and FSS. (Top) In MSN, colour stimulus information is transformed using colour target position into evidence for/against a response direction before being accumulated towards a response. (Bottom) In FSS, raw colour stimulus information is accumulated towards a categorial choice. This categorical choice must be transformed using colour target position and context into a response direction. Bold text marks the input variables used in Mante et al (top) and Flesch et al (bottom). The motion variable is matched between the two studies, whereas the colour variable is not.